\documentclass[a4paper,fleqn,usenatbib]{mnras}

\usepackage{mathptmx}
\usepackage{txfonts}

\usepackage[T1]{fontenc}
\usepackage{ae,aecompl}
\usepackage{multicol}
\usepackage{graphicx}
\usepackage{color}
\usepackage{graphicx}
\usepackage{amssymb}
\usepackage{rotating}
\usepackage{longtable}
\usepackage{lscape}

\newcommand{\figref}{Figure~\ref}

\title[The Sagittarius tidal stream around Whiting\,1]{The southern leading and trailing wraps of the Sagittarius tidal stream around the globular cluster Whiting\,1}

\author[J. A. Carballo-Bello et al.]{
J. A. Carballo-Bello,$^{1,2}$\thanks{E-mail: jcarballo@astro.puc.cl}, 
J. M. Corral-Santana$^{1,3}$, D. Mart\'inez-Delgado$^{4}$,\newauthor 
A. Sollima$^{5}$, R. R. Mu\~noz$^{6}$, P. C\^ot\'e$^{7}$, S. Duffau$^{2,1}$, M. Catelan$^{1,2}$, E. K. Grebel$^{4}$\\
\\
$^{1}$Instituto de Astrof\'isica, Facultad de F\'isica, Pontificia Universidad Cat\'olica de Chile, Av. Vicu\~na Mackenna, 4860, 782-0436,\\
Macul, Santiago, Chile\\
$^{2}$Millenium Institute of Astrophysics, Santiago, Chile\\
$^{3}$European Southern Observatory, Alonso de C\'ordova 3107, Casilla 19001, Santiago, Chile\\
$^{4}$Astronomisches Rechen-Institut, Zentrum f\"ur Astronomie der Universit\"at Heidelberg, M\"onchhofstr. 12-14, D-69120 Heidelberg, Germany\\
$^{5}$INAF Osservatorio Astronomico di Bologna, via Ranzani 1, I-40127 Bologna, Italy\\
$^{6}$Departamento de Astronom\'ia, Universidad de Chile, Camino El Observatorio 1515, Las Condes, Santiago, Chile\\
$^{7}$National Research Council of Canada, Herzberg Astronomy and Astrophysics, Victoria, BC, V9E 2E7, Canada\\
}

\date{Accepted XXX. Received YYY; in original form ZZZ}

\pubyear{2016}

\begin{document}
\label{firstpage}
\pagerange{\pageref{firstpage}--\pageref{lastpage}}
\maketitle

\begin{abstract}
We present a study of the kinematics of 101 stars observed with VIMOS around Whiting\,1, a globular cluster embedded in the Sagittarius tidal stream. The obtained velocity distribution shows the presence of two wraps of that halo substructure at the same heliocentric distance as that of the cluster and with well differentiated mean radial velocities. The most prominent velocity component seems to be associated with the trailing arm of Sagittarius with $<v_{\rm r}> \sim -130$\,km\,s$^{-1}$, which is consistent with the velocity of Whiting\,1. This result supports that this globular cluster was formed in Sagittarius and recently accreted by the Milky Way. The second component with $<v_{\rm r}> \sim 120$\,km\,s$^{-1}$ might correspond to the leading arm of Sagittarius, which has been predicted by numerical simulations but with no conclusive observational evidence of its existence presented so far. This detection of the old leading wrap of Sagittarius in the southern hemisphere may be used to confirm and further constrain the models for its orbit and evolution.
\end{abstract}


\section{Introduction}

  \begin{figure*}
     \begin{center}
      \includegraphics[scale=0.40]{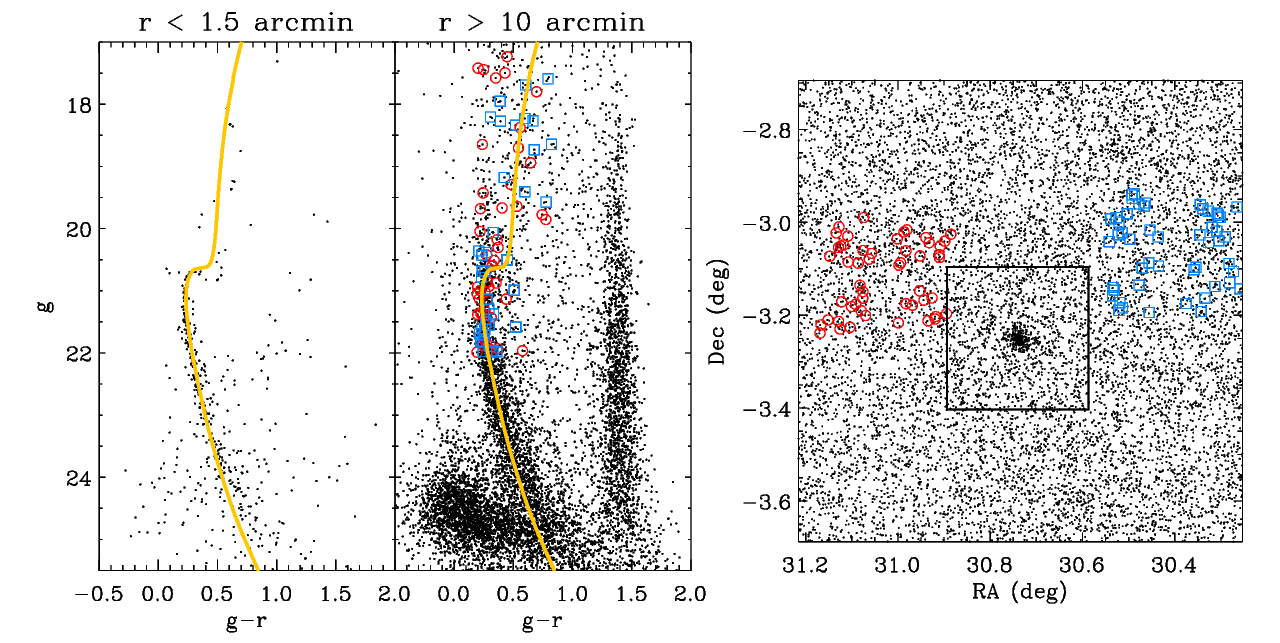}
      \caption[Observations]{\emph{Left and middle panels}: Megacam CMDs corresponding to the stars in the catalog at distances $r < 1.5$\,arcmin and $r > 10$\,arcmin from the center of Whiting\,1, respectively. The underlying MS observed in the middle panel is likely associated with the Sgr stream at the distance $d_{\odot} \sim 30$\,kpc as derived from isochrone fitting . The target stars in Fields 1 and 2 are displayed as red and blue circles, respectively. The yellow solid line corresponds to a $t \sim 10$\,Gyr and [Fe/H] $\sim -1.5$ population \emph{Right panel:} distribution of the target stars around Whiting\,1 using the same color code to identify the fields observed. The position of Whiting\,1 is indicated by a 10\,arcmin $\times$ 10\,arcmin black square.}
\label{observations_fig}
     \end{center}
   \end{figure*}

The accretion of massive satellites is believed to have contributed to the formation of a significant fraction of galaxy haloes by transferring a large amount of gas, stars and globular clusters (GC). Compelling evidence for the latter comes from the GC systems of nearby galaxies (e.g., M\,31), where a large fraction of their globulars seem to align with tidal streams \citep{Mackey2010,Mackey2013,Huxor2014,Veljanoski2014}. As for the Milky Way, the accretion of GCs in the hierarchical formation scheme has been long proposed because of the existence of at least two distinct subgroups in the Galactic halo \citep[e.g.][]{Searle1978,Zinn1993,Marin-Franch2009,Forbes2010,Leaman2013,Zaritsky2016}. 

The most pronounced accretion event in the Galaxy is the one corresponding to the Sagittarius \citep[Sgr;][]{Ibata1994} dwarf spheroidal and its associated stellar structure, the so-called Sgr tidal stream \citep[e.g.][]{Martinez-Delgado2001,Newberg2002,Majewski2003,Belokurov2006a,Koposov2012,Huxor2015}. Given the spatial extent of that system, it seems likely that some of the Galactic GCs might have formed within Sgr and been later accreted by the Milky Way. Indeed, four GCs are immersed in the Sgr main body \citep[M\,54, Arp\,2, Terzan\,7 and Terzan\,8;][]{DaCosta1995} and multiple Galactic halo GCs have been associated with the accreted galaxy \citep[e.g.][]{Dinescu2000,Bellazzini2002,Palma2002, Martinez-Delgado2002,Bellazzini2003,Carraro2009,Forbes2010,Dotter2011,Sbordone2015}. The total number of GCs likely associated with Sgr may reach $\sim 20\%$ of the outer halo GC population at $R_{\rm G} > 10$\,kpc \citep{Bellazzini2003}, and \cite{Law2010b} found 9 globulars compatible with the orbit of the stream predicted by their model \citep[][ hereafter LM10]{Law2010a}. The systematic search for tidal streams in wide fields around GCs by \citet[][hereafter CB14]{Carballo-Bello2014} revealed more clusters  surrounded by Sgr stars but often at different heliocentric distances. However, their search method cannot detect debris situated at closer distances in the plane of the Sgr orbit because of its lower surface brightness. A final conclusion about the extra-Galactic origin of these GCs requires the kinematic characterization both of the cluster and the underlying populations.
 
Whiting\,1 is a $6$\,Gyr old Galactic GC that has been associated with Sgr because of its projected position, distance ($d_{\odot} \sim 30$\,kpc) and radial velocity ($v_{\rm r} \sim -130$\,km\,s$^{-1}$), which are consistent with theoretical predictions for the stream in that region \citep{Carraro2007,Law2010a,Koposov2012,Valcheva2015}. \cite{Forbes2010} have shown that this cluster fits in the age-metallicity relation followed by other globulars associated with Sgr. In CB14, a prominent underlying population (likely Sgr-related) was unveiled around this cluster, at the same heliocentric distance and with the highest deviation from the synthetic Milky Way photometric models. In this work, we complete the study on the origin of Whiting\,1 by including kinematic information for a sample of stream stars around the cluster.

\section{Observations}

For the selection of targets for spectroscopy, we have used in this work the photometric catalogs generated in a Megacam$@$CFHT and Megacam$@$Magellan survey of all outer Galactic halo satellites (R. R. Mu\~noz et al., in preparation). \figref{observations_fig} shows the Megacam CMD corresponding to Whiting\,1 (left panel) and its surroundings (middle panel), where a well-populated main sequence (MS) in the latter confirms the presence of the Sgr tidal stream. The same isochrone from the \cite{Dotter2008} database used by CB14 ($t = 10$\,Gyr, ${\rm [Fe/H] } = -1.5$) is fitted to the Sgr population and we thus confirm that both cluster and stream are at a similar heliocentric distance of $d_{\odot} \sim $30\,kpc. We selected targets in a box with $0.2 < g-r < 0.8$ and $14.7 < g < 21.9$ for stars with distances greater than 10 times the \cite{King1962} tidal radius of Whiting\,1 \citep[$r_{\rm t} = 1$\,arcmin,][]{Carballo-Bello2012}, including Sgr turn-off stars and bright foreground Milky Way stars as control targets. Two fields around Whiting\,1 were selected for targeting based on the total number of slits placed when designing the masks, with a total number of 67 and 64 stars observed in the fields 1 and 2, respectively.   

Spectroscopic observations have been performed using the VIsible MultiObject Spectrograph (VIMOS) mounted at the 8.2\,m Very Large Telescope (Cerro Paranal, Chile) during 8 hours in service mode. The mid-resolution grism and the filter GG475 allowed a spectral coverage from 5000 to 8000\,\AA~ with a resolution $R = 580$.  Spectra are the result of combining 4 independent exposures of 1740\,s with an average signal-to-noise ratio of 30, which have been extracted using the \textsc{ESOREFLEX} pipeline for VIMOS. The instrumental flexure was estimated by measuring the position of a few sky lines which were used to correct the individual spectra.

Radial velocities were derived by cross-correlating our normalized spectra with a list of templates  smoothed to the resolution of our results. The sample of templates used was compiled by \cite{Pickles1998} and we only considered MS or subgiant stars ranging from O to M spectral types in this procedure. The final radial velocity value was adopted from the template providing the better correlation coefficient. Radial velocities have been then corrected for the Earth motion relative to the heliocentric rest by using the \textsc{IRAF} task rvcorrect. The average error found was $\sigma_{v_{\rm r}} \sim 30$\,km\,s$^{-1}$.

\section{Results and discussion}

  \begin{figure}
     \begin{center}
      \includegraphics[scale=0.24]{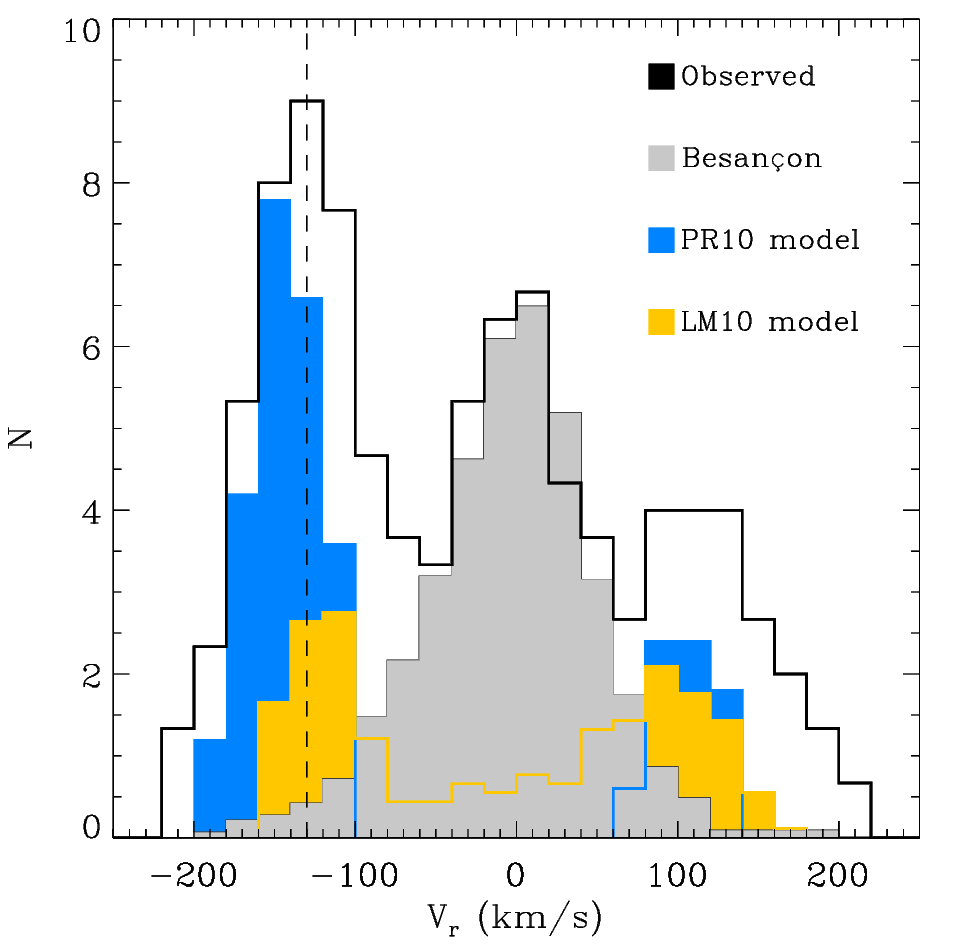}
      \vskip -0.1 cm
      \caption[Histogram]{Radial velocity distribution obtained for the sample of stars around Whiting\,1 (black solid line). The grey area indicates the velocity distribution as predicted by the Besan\c con model for the same line-of-sight, while the blue and yellow areas represent the expected velocities for Sgr stars, according to PR10 and LM10 models, respectively. Note that the grey area has been scaled to match the central peak while the blue and yellow areas have been arbitrarily scaled for comparison with the results. The dashed vertical line indicates the radial velocity of the GC.}
\label{histogram_fig}
     \end{center}

   \end{figure}

The derived radial velocity distribution of the 101 target stars is shown in \figref{histogram_fig}. The histogram is constructed using a bin size of 20\,km\,s$^{-1}$ and the observed morphology remains similar when the parameters used to generate it are changed or only one of the fields is considered. Three different components are clearly distinguished in the velocity distribution with approximate mean velocities of $v_{\rm r} = $-130,  0 and 120\,km\,s$^{-1}$ in descending order of number of stars.
 
The inclusion of bright foreground stars as targets ensures the presence of Milky Way stars along that line-of-sight. We evaluate such a contribution by comparing our results with the velocity distribution predicted by the Besan\c con synthetic model \citep{Robin2003}. We performed 10 simulations using the default parameters for the position in the sky of Whiting\,1 and for a solid angle of 1\,deg$^{2}$. We only considered those synthetic stars in the same color-magnitude range as of our targets, and used the same step and window width. The comparison histogram results from the co-addition of the 10 synthetic distributions, and it was scaled to fit its maximum value to our results (see \figref{histogram_fig}). This procedure confirms that the bulk of stars at $v_{\rm r} \sim 0$\,km\,s$^{-1}$ is mostly composed of Milky Way stars, while the peaks at $v_{\rm r} = -130$ and 120\,km\,s$^{-1}$ represent excesses of stars not associated with any known Galactic population. 

%\footnote{Website: http://model.obs-besancon.fr/} 

We derive velocity distributions from the \cite[][hereafter PR10]{Penarrubia2010} and LM10 models of the Sgr tidal stream along the line-of-sight of Whiting\,1. We have selected in both cases all those model stars in an area of 2\,deg~$\times$~2\,deg around the cluster position and generated a histogram using the same bin and window sizes used to build the previous ones. We then arbitrarily scaled the distributions and overplotted the results in \figref{histogram_fig}. Both PR10 and LM10 predict 2 components of Sgr in that area of the sky with mean velocities $<v_{\rm r}> \sim -130$ and $\sim$ 115\,km\,s$^{-1}$. The position of these peaks are compatible with the ones derived from the velocity distribution of the target stars. According to the classification provided by these models, the bulk of Sgr stars around $v_{\rm r} \sim -130$\,km\,s$^{-1}$ might belong to the trailing arm of that halo substructure while the less significant peak might correspond to the leading arm. The mean heliocentric distance predicted by PR10 for both wraps in this area of the sky is $d_{\odot} \sim 30$\,kpc. On the other hand, LM10 predicts mean distances of $d_{\odot} \sim 21$\,kpc and $\sim28$\,kpc for the leading and trailing arms, respectively.

\begin{figure}
     \begin{center}
          \vskip -0.3 cm
      \includegraphics[scale=0.3]{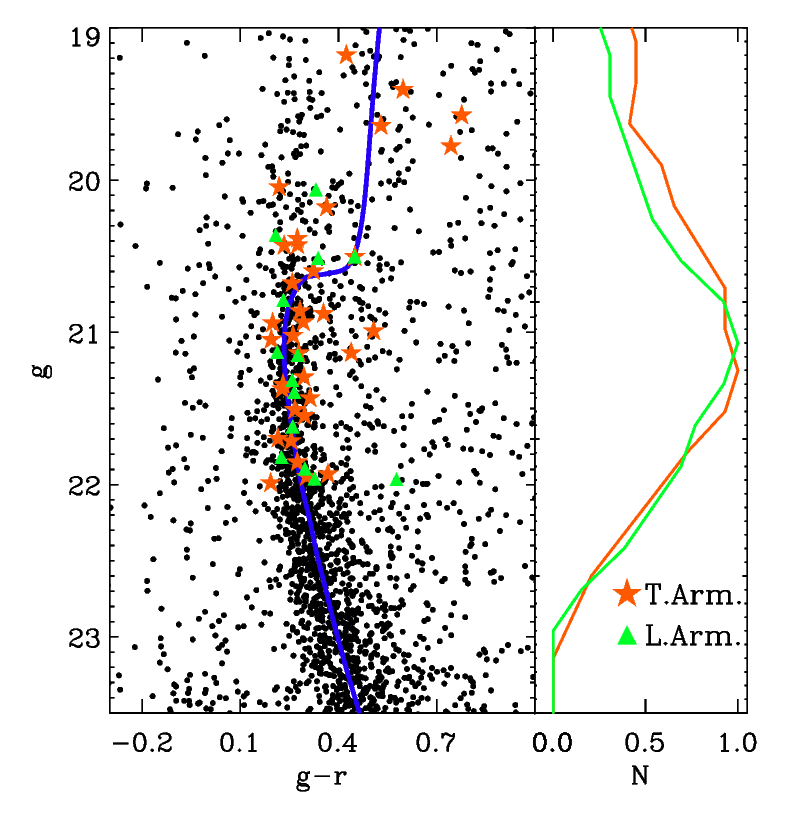}
      \caption[cmd2]{\emph{Left:} CMD for stars beyond 10\,arcmin from the center of Whiting\,1. The positions in the diagram of stars belonging to the Sgr trailing and leading wraps are indicated as orange stars and green triangles, respectively. The blue isochrone corresponds to a $t \sim 10$\,Gyr and [Fe/H] $\sim -1.5$ population. \emph{Right:} normalized distribution of  stars belonging to both arms as a function of $g$ in the range $19 < g < 23.5$ and following the same color code as in the left panel.}
\label{cmd2_fig}
     \end{center}
   \end{figure}

The expected two sections of the Sgr tidal stream along the line-of-sight to Whiting\,1 are evident in our spectroscopic results but not in the CMD shown in \figref{observations_fig}. We analyze the distribution of target stars belonging to the Sgr components as a function of $g$ by counting stars associated with the leading ($-160 < v_{\rm r}$ [\,km\,s$^{-1}$]$ < -100$) and trailing ($90 < v_{\rm r}$ [\,km\,s$^{-1}$]$ < 150$) arms with a bin size of $\delta\,g = 0.1$. \figref{cmd2_fig} shows the normalized distribution obtained for both populations and we are not able to determine a shift in the $g$ band between the subsamples of target stars. This might indicate that the leading and trailing arms of Sgr are spatially coincident or have slightly different distances, making it difficult to distinguish the populations in the CMD. Therefore, our results are in good agreement with the predictions made by the PR10 model both for the position and the kinematics of the leading and trailing components of the Sgr tidal stream around Whiting\,1.    

According to the PR10 classification, target stars with radial velocities around $v_{\rm r} \sim 120$\,km\,s$^{-1}$ might be associated with an \emph{old} component of the leading arm of Sagittarius, accreted a long time ago ($> 2$\,Gyr). While the northern and southern trailing arms have been well characterized using different tracers such as RR\,Lyrae, blue horizontal branch, sub-giant branch and red clump stars \citep[e.g.][]{Koposov2012,Belokurov2014}, the southern leading arm predicted both by PR10 and LM10, has remained undetected. This might be due to the coincidence of its projected path with that of the trailing arm or to its lower surface-brightness. Only tentative detections have been reported using alternative tracers such as carbon stars, with distances and velocities consistent with our results along that line-of-sight \citep[][]{Huxor2015}.

Alternatively, the group of stars with positive radial velocities might belong to a different halo substructure. The Cetus Polar stream \citep{Newberg2009,Yam2013} is located at a compatible heliocentric distance of $d_{\odot} \sim $33\,kpc at its nearest point to Whiting\,1. However, that substructure is crossing the sky at $\ell \sim$ 140$^{\circ}$ and presents a very low density of member stars to become such an important contribution in our velocity distribution. Additional streams or overdensities have not been reported in this line-of-sight, and  so the Sgr tidal stream remains as the likely underlying population. This first spectroscopic detection of the leading arm of that tidal stream confirms one of the predictions made by the available Sgr models and could be used in the future to better constrain the complex orbit of the stream.

\begin{figure}
     \begin{center}
      \includegraphics[scale=0.26]{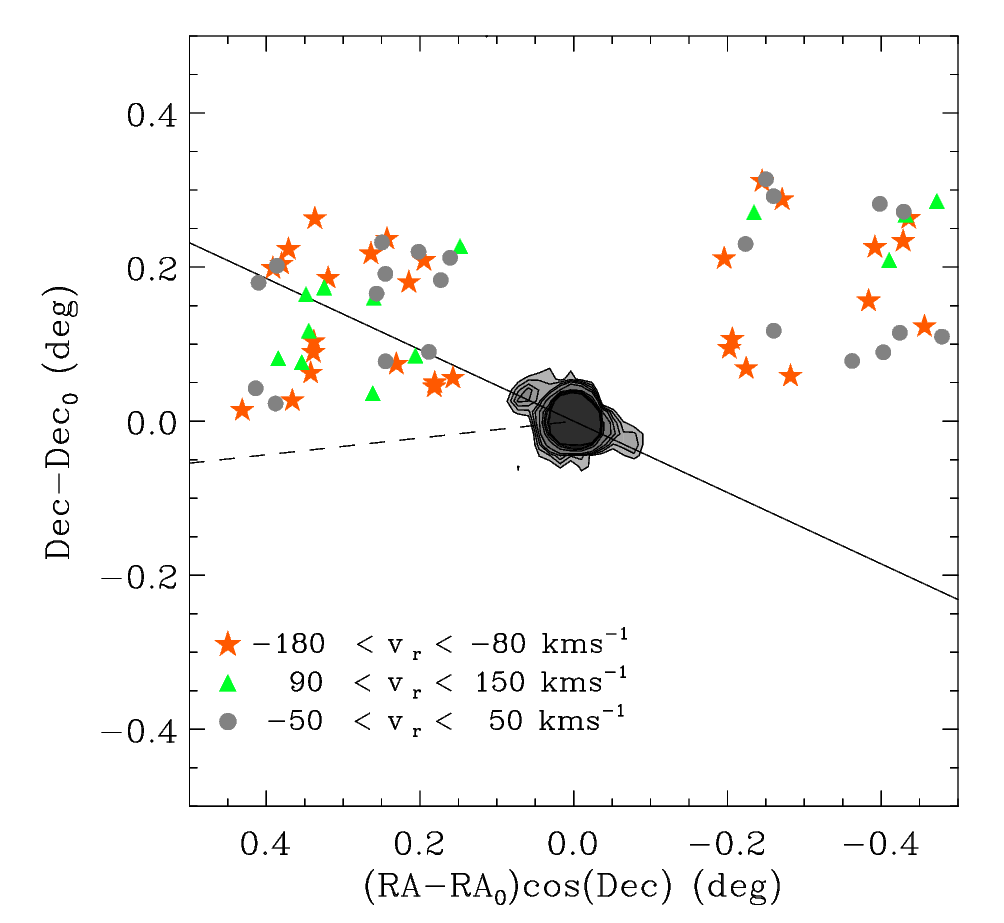}

      \caption[Map]{Density map generated for an area of 1\,deg $\times$ 1\,deg centered in the coordinates of Whiting\,1 [RA$_{0}$,Dec$_{0}$]=[30.73$^{\circ}$,-03.25$^{\circ}$]. Isodensity lines are drawn at levels of 3, 3.5, 4, 5, 7.5, 9, 10 and > 12 times the mean background level. The solid line indicates the mean orbit of the Sgr tidal stream along that line-of-sight according to PR10, whereas the dashed line indicates the direction of the Milky Way center. The symbols indicating the position of target stars are color-coded according to the velocity ranges indicated in the legend.}
\label{densitymap_fig}
     \end{center}
   \end{figure}

Once we have established the nature of the underlying population around Whiting\,1, we include in our analysis the radial velocity measured for this cluster, set at $v_{\rm r} \sim -130.6 \pm 1.8$\,km\,s$^{-1}$ \citep{Carraro2007} . The distance and velocity of the trailing arm of Sgr is compatible with that of Whiting\,1 (see \figref{histogram_fig}). According to PR10 model, that wrap is composed of stars stripped away during the last $\sim 0.6$\,Gyr, thus Whiting\,1 was accreted by the Milky Way during the last passage of Sgr. \cite{Law2010b} proposed 5 GCs with a higher probability of belonging to the Sgr GC system (Arp\,2, M\,54, NGC\,5634, Terzan\,8 and Whiting\,1) while other 4 clusters (Berkeley\,29, NGC\,5053, Pal\,12 and Terzan\,7) display a lower but still significative probability. Among the most likely members, only NGC\,5634 and Whiting\,1 are far away from the Sgr core and both are associated with the trailing arm in the LM10 model. Although no underlying population was found at a compatible distance around NGC\,5634 in CB14, the abundances derived by \cite{Sbordone2015} for this cluster are consistent with those of the main body of Sgr. \cite{Law2010b} also show that Whiting\,1 is the second closest GC to the plane containing the orbit of Sgr, after M\,54. Our observations support the scenario in which Whiting\,1 was formed in the Sgr dwarf spheroidal and later deposited in the Galactic halo. 

The minimum distances of the target stars from the center of Whiting\,1 are 10 and 13\, arcmin in fields 1 and 2, respectively, which ensure a low probability of including cluster stars. This is confirmed by the distribution of trailing and leading arm stars shown in \figref{densitymap_fig}, where none of the populations seem to be clustered around Whiting\,1. We also investigate the structure of the cluster searching for the signature of tidal tails in the CFHT photometry. We perform a matched-filter analysis of the distribution of stars associated with Whiting\,1 following the procedure described in \cite{Rockosi2002}. The density contours derived for the GC are shown in \figref{densitymap_fig}, where the position of the target stars are also overplotted. Given the relative position of spectroscopic targets and cluster, we discard the presence of cluster members in our sample as the origin of the two-peak distribution observed. Interestingly, the cluster has a hint of elongations in opposite directions suggesting that the cluster has suffered severe tidal stripping. This was already suggested by the deviation from a King profile in the outermost parts of the cluster observed by \cite{Carraro2007}. We have obtained a mean orbit of the Sgr tidal stream according to the PR10 predictions for the range 130$^{\circ} < \ell < 190^{\circ}$. That mean orbit is overplotted in \figref{densitymap_fig} and the elongations emerging from the cluster seem to be aligned with the orbit of the stream. We interpret this result as new evidence of the likely association of Whiting\,1 with the Sgr tidal stream.

\section{Conclusions}

We have used VIMOS to derive radial velocities for a sample of 101 potential Sagittarius tidal stream members around the GC Whiting\,1. Our velocity distribution shows the presence of two populations with mean radial velocities, which are consistent with the predictions made by numerical simulations for the stream in this area of the sky. Both components are found at the same heliocentric distance that of the cluster.

The most prominent peak with $<v_{\rm r}> \sim -130$ is composed of stars  associated with the trailing arm of Sagittarius. The radial velocity of Whiting\,1, compatible with that of this wrap, and  the elongation of the stellar density isopleth of cluster stars suggests that this GC was recently accreted by the Milky Way from the Sagittarius dwarf galaxy. The second component in the radial velocity distribution with $<v_{\rm r}> \sim 120$\,km\,s$^{-1}$ seems to be associated with the leading arm of the Sagittarius tidal stream. This section of the stream, accreted a long time ago (> 2\,Gyr), is predicted by all the available models for Sagittarius but no observational evidence has been reported to confirm its existence. This first detection of the leading arm of Sagittarius in the southern hemisphere will help future simulations to describe the overall structure of the tidal stream and impose new con-
straints to its orbit and the Milky Way accretion history.

\section*{Acknowledgements}

We warmly thank the anonymous referee for his/her helpful comments and suggestions. We acknowledge financial support to CONICYT-Chile FONDECYT Postdoctoral Fellowships 3160502 (JAC-B) and 3140310 (JMC-S). JAC-B and MC received support from the Ministry for the Economy, Development, and Tourism's Programa Iniciativa Cient\'ifica Milenio through grant IC120009, awarded to the Millennium Institute of Astrophysics (MAS) and from CONICYT's PCI program through grant DPI20140066. DMD and EKG acknowledge funding from Sonderforschungsbereich SFB 881 ''The Milky Way System" (subproject A2) of the German Research Foundation (DFG). Based on data products from observations made with ESO Telescopes at the La Silla Paranal Observatory under ESO programme ID 091.D-0446(D). Partially based on observations obtained at the CFHT, which is operated by the National Research Council of Canada, the Institut National des Sciences de l\' \,Univers of the Centre National de la Recherche Scientifique of France, and the University of Hawaii. Thanks to M. D. Mora for his help with the radial velocitiy determination and to V. Belokurov and S. Koposov for their useful comments.

\def\jnl@style{\it}                       % Defines journal style in italics
\def\mnref@jnl#1{{\jnl@style#1}}          % Defines \mnref command to call journals
\def\aj{\mnref@jnl{AJ}}                   % Astronomical Journal
\def\apj{\mnref@jnl{ApJ}}                 % Astrophysical Journal
\def\apjl{\mnref@jnl{ApJL}}               % Astrophysical Journal, Letters
\def\aap{\mnref@jnl{A\&A}}                % Astronomy and Astrophysics
\def\mnras{\mnref@jnl{MNRAS}}             % Monthly Notices of the RAS
\def\nat{\mnref@jnl{Nat.}}                % Nature
\def\iaucirc{\mnref@jnl{IAU~Circ.}}       % IAU Circulars
\def\atel{\mnref@jnl{ATel}}               % Astronomers Telegram
\def\iausymp{\mnref@jnl{IAU~Symp.}}       % IAU Symposium
\def\pasp{\mnref@jnl{PASP}}               % Astronomical Journal
\def\araa{\mnref@jnl{ARA\&A}}             % Astronomical Journal
\def\apjs{\mnref@jnl{ApJS}}               % Astronomical Journal
\def\aapr{\mnref@jnl{A\&A Rev.}}          % Astronomical Journal

\bibliographystyle{mn2e}
\bibliography{biblio}

% Don't change these lines
\bsp	% typesetting comment
\label{lastpage}
\end{document}